\documentclass[preprint,10pt,review,times]{elsarticle}
\usepackage{amsmath}
\usepackage{amssymb}
\usepackage{amsthm}
\usepackage{lineno}
\usepackage{caption}
\usepackage{subcaption}

\journal{Journal of Molecular Spectroscopy}

\begin{document}
\begin{frontmatter}
\title{Ghost features in Doppler-broadened spectra of rovibrational transitions in trapped HD$^+$ ions}
\author[LLB]{Sayan Patra}
\author[LLB]{J.C.J.~Koelemeij\corref{cor1}}
\ead{j.c.j.koelemeij@vu.nl}
\cortext[cor1]{Corresponding author}
\address[LLB]{LaserLaB and Department of Physics and Astronomy, Vrije Universiteit Amsterdam,\\
De Boelelaan 1081, 1081 HV, Amsterdam, The Netherlands}

\begin{abstract}
Doppler broadening plays an important role in laser rovibrational spectroscopy of trapped deuterated molecular hydrogen ions (HD$^+$), even at the millikelvin temperatures achieved through sympathetic cooling by laser-cooled beryllium ions. Recently, Biesheuvel \textit{et al.} \cite{Biesheuvel2016} presented a theoretical lineshape model for such transitions which not only considers linestrengths and Doppler broadening, but also the finite sample size and population redistribution by blackbody radiation, which are important in view of the long storage and probe times achievable in ion traps.  Here, we employ the rate equation model developed by Biesheuvel \textit{et al.} to theoretically study the Doppler-broadened hyperfine structure of the $(v,L):(0,3)\rightarrow(4,2)$ rovibrational transition in HD$^+$ at 1442~nm. We observe prominent yet hitherto unrecognized ghost features in the simulated spectrum, whose positions depend on the Doppler width, transition rates, and saturation levels of the hyperfine components addressed by the laser. We explain the origin and behavior of such features, and we provide a simple quantitative guideline to assess whether ghost features may appear. As such ghost features may be common to saturated Doppler-broadened spectra of rotational and vibrational transitions in trapped ions composed of partly overlapping lines, our work illustrates the necessity to use lineshape models that take into account all the relevant physics.
\end{abstract}
\end{frontmatter}

\section{Introduction} \label{sec:intro}
Molecular hydrogen ions are the simplest molecules in nature, and their simple three-body structure allows highly precise calculations of their level structure and their interaction with electromagnetic fields. Vibrational level energies in the electronic ground state have been calculated with a relative uncertainty of the order of $10^{-11}$ \cite{Korobov2014a,Korobov2014b}, and also the theoretical hyperfine structure is well known \cite{Bakalov2006,Korobov2016}. In the deuterated molecular hydrogen ion, HD$^+$, rovibrational overtone transitions in the (near-)infrared are weakly dipole allowed, leading to natural linewidths of the order of 10 Hz. In combination with the availability of high-accuracy theory, this has stimulated high-resolution laser spectroscopy of HD$^+$ in ion traps \cite{Roth2006,Koelemeij2007b,Bressel2012,Zhong2015,Biesheuvel2016}. Typically, the transitions are detected through $(1+1')$ resonance-enhanced multiphoton dissociation (REMPD) by monitoring the loss of HD$^+$ ions from the trap for various frequencies of the first (probe) laser, near or at resonance. To reduce Doppler broadening, HD$^+$ ions are sympathetically cooled by laser-cooled Be$^+$ ions to temperatures of about 10 mK~\cite{Blythe2005}.  Typically, the observed optical spectra have an irregular shape, which is caused by several partly overlapping Doppler-broadened hyperfine components, and effects of saturation (due to depletion of the finite amount of HD$^+$ available in the trap) and population redistribution among rotational levels by blackbody radiation (BBR). In early lineshape models, the contributions of individual hyperfine components were simply summed \cite{SpezeskiThesis,Roth2006,Koelemeij2007a,Bressel2012}. However, Biesheuvel \textit{et al.} presented an extended lineshape model which also accounted for the finite HD$^+$ sample size, the actual dissociation rate, and population redistribution by BBR. This model proved essential for performing the most precise vibrational frequency measurement in a molecular ion so far, leading to the most stringent test of molecular theory to date,  improved bounds on the existence of possible `new physics', and the first determination of the proton-electron mass ratio from molecular spectroscopy~\cite{Biesheuvel2016}.

Here, we theoretically study the Doppler-broadened hyperfine structure of the $(v,L):(0,3)\rightarrow(4,2)$ rovibrational transition in HD$^+$. This transition is the first step of the two-photon transition proposed by Tran \textit{et al.} \cite{Tran2013}, which should enable Doppler-free spectroscopy of HD$^+$ at improved resolution. Apart from the Doppler-free signal, a background signal due to Doppler-broadened excitation will be present \cite{Tran2013,Karr2016}. To assess the effect of the latter, we employ the model of Biesheuvel \textit{et al.} to study the $(v,L):(0,3)\rightarrow(4,2)$ transition. We find that depending on the conditions (HD$^+$ temperature, transition rates, BBR temperature, and saturation level) this spectrum contains several ghost features, \textit{i.e.} relatively narrow peaks in parts of the spectrum where no signal is expected as there are no nearby hyperfine components. In this work we describe these ghost features in detail and explain their origin, leading us to conclude that such features may be common to any saturated Doppler-broadened spectroscopy of transitions involving partly overlapping (hyperfine) subcomponents.

This article is organized as follows. In Sec.~\ref{sec:rateeqreview}, we briefly review the rate equation model that was developed by Biesheuvel \textit{et al.}, followed by the calculation of the $(v,L):(0,3)\rightarrow(4,2)$ spectrum in Sec.~\ref{sec:calcspec}. The ghost features are described in detail in Sec.~\ref{sec:ghostfeatures}, followed by a description of their possible impact in Sec.~\ref{sec:impact} and the conclusions in Sec.~\ref{sec:summary}.

\section{Brief review of the rate equation model} \label{sec:rateeqreview}
\begin{figure}[!ht]
\centering
\includegraphics[width=1.0\linewidth]{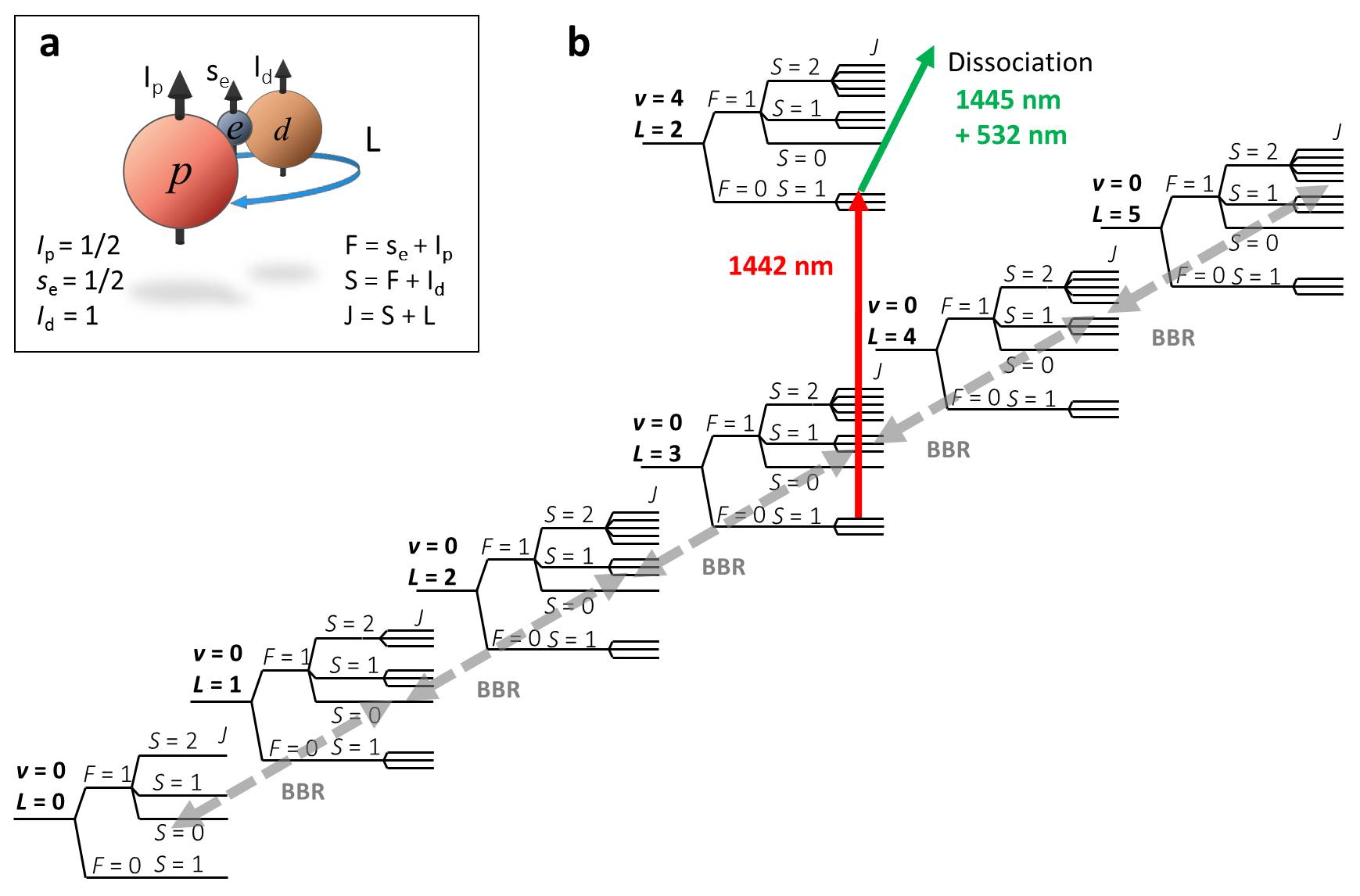}
\caption{Partial level scheme (not to scale) of the relevant rovibrational and hyperfine levels of HD$^+$ (all in the 1s$\sigma$ ground electronic state), and a schematic representation of the REMPD process. (a) Angular momentum picture of the HD$^+$ molecular ion and the angular momentum coupling scheme used. (b) Rotational levels in $v=0$ are coupled by BBR and spontaneous emission (gray dashed arrows). The $v=0,L=3$ state is probed by the 1442~nm laser, which is subsequently dissociated following a second (resonant) 1445~nm step to $v=9,L=3$ and a third 532~nm step which promotes the molecule to the predissociating 2p$\sigma$ electronic state. Although this scheme actually corresponds to $(1+1'+1'')$ REMPD, we treat the final two steps as a single dissociative transition (\textit{i.e.} $(1+1')$ REMPD).}
\label{fig:levelscheme}
\end{figure}
The idea here is to describe the sample of HD$^+$ ions by a state vector, $\rho(t)$, describing the population of all 62 hyperfine levels in the $(v=0,L\leq5)$ manifold (Fig.~\ref{fig:levelscheme}), which contains 97.6\% of the population at a BBR temperature of 300~K. Ignoring states with $L>5$ minimizes the computational resources needed for the numerical calculations. A measure of the total number of HD$^+$ ions can be obtained by summing over all elements of $\rho(t)$, from which we can also obtain the spectroscopic REMPD signal (\textit{i.e.} the fractional loss of HD$^+$ due to REMPD). The hyperfine structure arises from magnetic interactions between the proton spin, electron spin, deuteron spin and the molecular rotation, and is described in detail in references \cite{SpezeskiThesis,Bakalov2006}. States with $L=0$, $L=1$ and $L\geq 2$ possess 4, 10 and 12 hyperfine levels, respectively, each having its own total angular momentum, $J$.  We ignore Zeeman splittings and shifts due to typical magnetic fields of $\sim 0.2$~mT as these are much smaller than the Doppler width of the transition at the 10~mK translational temperature of the HD$^+$ ions. During REMPD, the ions interact both with the probe radiation and the BBR field, and the time evolution of the population in the $v=0$ manifold is given by the time evolution of the state vector $\rho(t)$ as below,
\begin{equation} \label{eq:rateeq}
\frac{d}{dt}\rho(t)=M_\text{REMPD}\cdot \rho(t)+M_\text{BBR}\cdot \rho(t).
\end{equation}
Here, $M_\text{REMPD}$ and $M_\text{BBR}$ are matrices describing the interactions with the spectroscopy probe laser (which induces the rovibrational transition of interest) and the BBR field, respectively. Assuming that the ions which make the transition are instantly dissociated by the dissociation laser allows treating the interaction with the probe radiation as a simple loss process. The matrix elements of $M_\text{REMPD}$ are strictly diagonal and can be written as
\begin{equation}\label{eq:Mrempd}
M_{\text{REMPD},{\alpha \alpha}}= -\sum_{\alpha'}B_{\alpha \alpha'}D_z \left( \omega-\omega_{\alpha \alpha'},T_{\text{HD}^+} \right) \frac{I_\text{probe}}{c},
\end{equation}
where $B_{\alpha \alpha'}$ is the Einstein coefficient for stimulated absorption between the lower (hyperfine) level $\alpha$ in $v=0,L=3$ and the upper (hyperfine) level $\alpha'$ in $v=4,L=2$ with a transition frequency of $\omega_{\alpha \alpha'}$. The summation runs over all upper hyperfine states $\alpha'$ within the target rovibrational state. $D_z$  is a normalized  response function (typically a Voigt profile), which involves an integration over all probe laser frequencies observed in the frame of the HD$^+$ ions moving with thermally distributed velocities $v_z$ along the z-direction (which coincides with the symmetry axis of the linear Paul trap assumed here, and also with the direction of the wave vector of the cooling laser for Be$^+$ ions). For a thermal velocity distribution characterized by typical translational (secular) HD$^+$ temperatures $T_{\text{HD}^+}$ of several millikelvins, the Doppler width [MHz-range, see Fig.~\ref{fig:5KBBR}(b)] is much larger than the natural linewidth ($\sim 10$~Hz), so that $D_z$ is well described by a Doppler-broadened Gaussian lineshape function. Furthermore, $\omega \equiv 2\pi \nu$ is the angular frequency of the 1442~nm probe laser, $I_\text{probe}$ is the intensity of the probe radiation, and $c$ is the speed of light in vacuum.
%
%The matrix elements of  $M_\text{BBR}$ can be represented as
%\begin{equation}
%M_\text{BBR}=A_{ij}+(B_{ij}+B_{ji})W(\omega,T_\text{BBR})
%\end{equation}
%$A_{ij}$, $B_{ij}$ and $B_{ji}$ are the Einstein coefficients of spontaneous emission, stimulated emission and %stimulated absorption respectively.
%

The matrix $M_{\text{BBR}}$ contains both diagonal and off-diagonal elements, which take into account the rate of exchange of population between all involved levels $\alpha$ and $\alpha'$ mediated through all possible electric-dipole transitions,
\begin{equation*}
\begin{split}
&A_{\alpha' \alpha}, \\
&B_{\alpha' \alpha}W(\omega_{\alpha \alpha'},T_{\text{BBR}}), \\
&B_{\alpha \alpha'}W(\omega_{\alpha \alpha'},T_{\text{BBR}}),
\end{split}
\end{equation*}
which correspond to spontaneous emission, stimulated emission and absorption by BBR, respectively. $W(\omega,T_\text{BBR})$ is the energy density for BBR at frequency $\omega$ and temperature $T_\text{BBR}$ and can be expressed as
\begin{equation}
W(\omega,T_\text{BBR})=\frac{\hslash\omega^3}{\pi^2 c^3} \left( \text{e}^\frac{\hslash\omega}{k_{B}T_\text{BBR}}-1 \right)^{-1},
\end{equation}
where $\hslash$ and $k_{B}$ are the reduced Planck constant and Boltzmann constant, respectively. We will assume $T_\text{BBR}=300$~K unless noted otherwise.

In the following section, we briefly describe the steps involved in the calculation of the $(v,L):(0,3)\rightarrow(4,2)$ spectrum in HD$^+$.
\section{Calculation of the $(v,L):(0,3)\rightarrow(4,2)$ spectrum} \label{sec:calcspec}
The internal degrees of freedom of the HD$^+$ molecule (rotations and vibrations) have frequencies greater than 1~THz while the external degrees of freedom have frequencies less than 1~MHz. Hence, laser cooling the external degrees of freedom of the molecule effectively does not affect the internal degrees of freedom. However, the molecular rotation interacts relatively strongly with the ambient BBR field, leading to a thermal distribution of rotational-state population corresponding to 300~K~\cite{Koelemeij2007b}.%PhysRevA.

Since the molecules which have made the transition are efficiently dissociated, the probability of them decaying spontaneously to lower levels is negligible~\cite{Biesheuvel2016}. Hence, for the BBR interaction, only the $v=0$ manifold is considered. The $A_{\alpha \alpha'}$, $B_{\alpha \alpha'}$ and $B_{\alpha' \alpha}$ coefficients between lower and upper levels $\alpha$ and $\alpha'$ are calculated using the radial dipole matrix elements obtained in Ref. \cite{Koelemeij2011}. The hyperfine linestrengths are calculated following the approach of Refs. \cite{SpezeskiThesis,Bakalov2011}, and the corresponding Einstein rate coefficients are given in Ref.~\cite{Tran2013}. A manuscript giving a more detailed account of the lineshape model of Biesheuvel \textit{et al.} is currently in preparation \cite{Biesheuvel2016b}. Inserting these coefficients in Eq.~(\ref{eq:rateeq}) allows quantifying the saturation and the Doppler broadening effects. To obtain highly accurate lineshapes it is in principle also required to take into account other systematic effects, such as possible non-thermal velocity distributions of the HD$^+$ ions, Zeeman splitting, and various line shifts, one of the most significant causes being the AC Stark effect~\cite{Biesheuvel2016}. However, such effects are marginal compared to the effects of Doppler broadening, REMPD and BBR, and for the purpose of the study in this article only the REMPD and the BBR interactions with a thermal ensemble of HD$^+$ are considered.
\begin{figure}[!htb]
\centering
\includegraphics[width=0.75\linewidth]{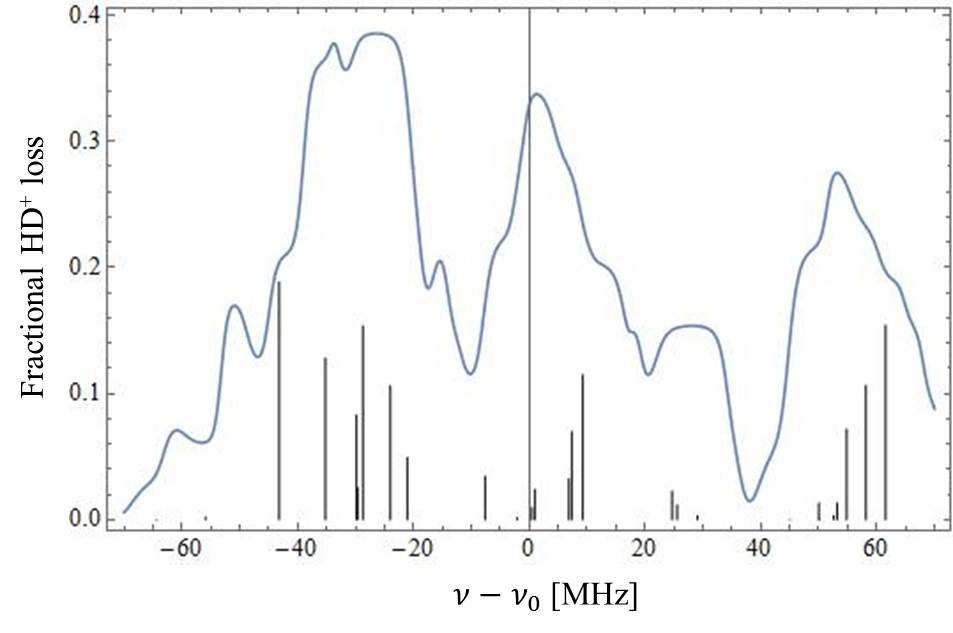}
\caption{Doppler-broadened REMPD spectrum of the $(v,L):(0,3)\rightarrow(4,2)$ transition for an ion temperature of 5~mK, laser power of 5~mW focussed in a waist of 100 $\mu$m, and a REMPD duration of 10~s (unless mentioned otherwise, these parameters are used for all figures). $\nu-\nu_0$ denotes the frequency of the probe laser with respect to the `hyperfine-less' rovibrational transition frequency $\nu_0$. The positions and relative strengths (in arbitrary units) of the underlying hyperfine components are indicated by the black sticks, which in some cases are barely visible (for example the one at $-64.3$~MHz). Ghost features can be seen at $\nu-\nu_0=-61$~MHz, $-51$~ MHz, and $-15$~MHz. There are other such features in the spectrum, but these are not visible clearly as they are encapsulated by the broader Doppler profile.}
\label{fig:fullspectrum}
\end{figure}

Assuming a probe duration of 10~s, the rate equations are solved using the algorithm \verb+NDSolve+ of Mathematica, resulting into  simulated spectra as plotted in Fig. \ref{fig:fullspectrum}.
In Fig.~\ref{fig:fullspectrum}, at frequencies of $\nu-\nu_0=-61$~MHz, $-51$~MHz and $-15$~MHz, relatively narrow peaks can be seen which do not correspond to any hyperfine component. Such ghost features turn out to be a common phenomenon in the simulated spectra, and we describe their origin and characteristics in more detail in the following section.

\section{Ghost features in $(v,L):(0,3)\rightarrow(4,2)$ single-photon spectrum} \label{sec:ghostfeatures}
%In this section, we analyze and explain the origin of these ghost features as seen in Figure %\ref{fig:fullspectrum}.
\begin{figure}[h!]
\centering
\includegraphics[width=1.0\linewidth]{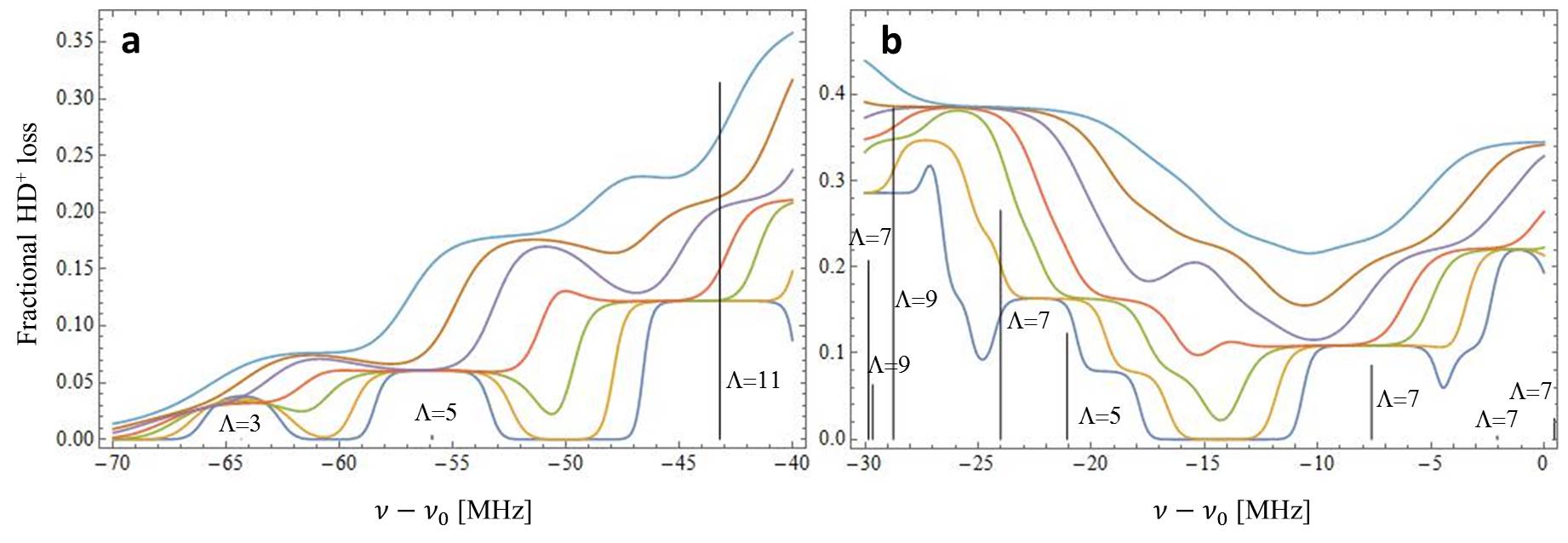}
\caption{REMPD signal plotted against frequency $\nu-\nu_0$ for different translational temperatures of the HD$^+$ ions. The hyperfine components are indicated by black sticks. From top to bottom, the curves correspond to ion temperatures 10~mK, 7~mK, 5~mK, 3~mK, 2~mK, 1~mK and 0.5~mK, respectively. Ghost features appear for increasing temperatures near $-50$ MHz and $-61$ MHz (a), and near $-15$ MHz (b).}
\label{fig:temp}
\end{figure}
\begin{figure}[!ht]
\centering
\includegraphics[width=0.8\linewidth]{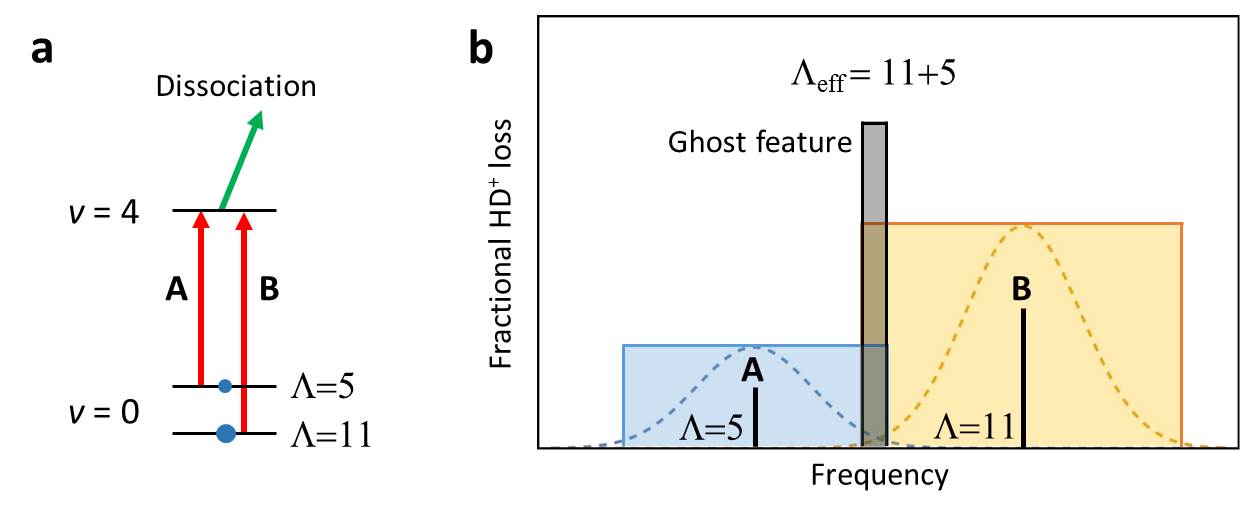}
\caption{Qualitative explanation of the origin of ghost features in saturated spectra consisting of partly overlapping lines. (a) Simplified three-level model of the $v=0 \rightarrow v=4$ transition in HD$^+$, characterized by two individual transitions, A and B, with different frequencies, linestrengths and lower-state multiplicities ($\Lambda=5$ and $\Lambda=11$, respectively). Excitation through transitions A and B is induced by a first laser, while the upper level is dissociated by a second laser to generate a signal (in the form of loss of HD$^+$). The dashed curves in the main plot indicate the Doppler-broadened response function $D_z$ for each line [Eq.~(\ref{eq:Mrempd})]. Clipping due to saturation is here exaggerated for clarity, and occurs at a signal level proportional to the multiplicity of the lower states addressed by the spectroscopy and dissociation lasers. In the region where the two clipped lineshapes overlap, the REMPD lasers address a larger population of HD$^+$ (proportional to the `effective' multiplicity $\Lambda_\text{eff}$), leading to enhanced signals and seemingly narrow ghost features.}
\label{fig:qualplot}
\end{figure}

Figure \ref{fig:temp}(a) zooms in on the spectrum of Fig.~\ref{fig:fullspectrum} in the frequency range ($-70$~MHz, $-40$~MHz). If the temperature of the ions is comparatively low (0.5 to 2~mK) such that the Doppler-broadened profiles of individual hyperfine components do not significantly overlap, only `true' hyperfine lines are visible, while ghost features are completely absent. Note that in our simulations the individual hyperfine lines are heavily saturated near their line centers, as evidenced by the flattened lineshapes. While each hyperfine component has it own transition strength (the relative transition strength is indicated by the size of the stick in the hyperfine stick spectrum of Figs.~\ref{fig:fullspectrum} and \ref{fig:temp}), the clipping due to saturation occurs at a signal level which is determined largely by the HD$^+$ population available in the initial hyperfine state. Within the hyperfine manifold of a given rovibrational level, this population is essentially equal to the multiplicity, $\Lambda \equiv 2J+1$, of that hyperfine state. If we now increase the temperature of the ion ensemble to 3~mK, the tails of the Doppler-broadened profiles of adjacent hyperfine components start to overlap, leading to additional signal in the region in between the adjacent hyperfine components. In this region, the laser addresses both adjacent hyperfine states simultaneously, and for sufficiently high saturation levels, the signal can rise \textit{above} the saturation levels of the individual hyperfine lines, thus giving rise to ghost features (Fig.~\ref{fig:qualplot}). The strength of these peaks is proportional to the sum of the lower-state multiplicities of the hyperfine lines between which they show up. This is explained qualitatively in Fig.~\ref{fig:qualplot}. Since these features are not necessarily saturated themselves, they may appear narrow as compared to the saturated, flattened hyperfine lines themselves. If not properly recognized, such ghost features may therefore falsely be identified as `true' hyperfine lines.

\begin{figure}[!ht]
\centering
\includegraphics[width=0.75\linewidth]{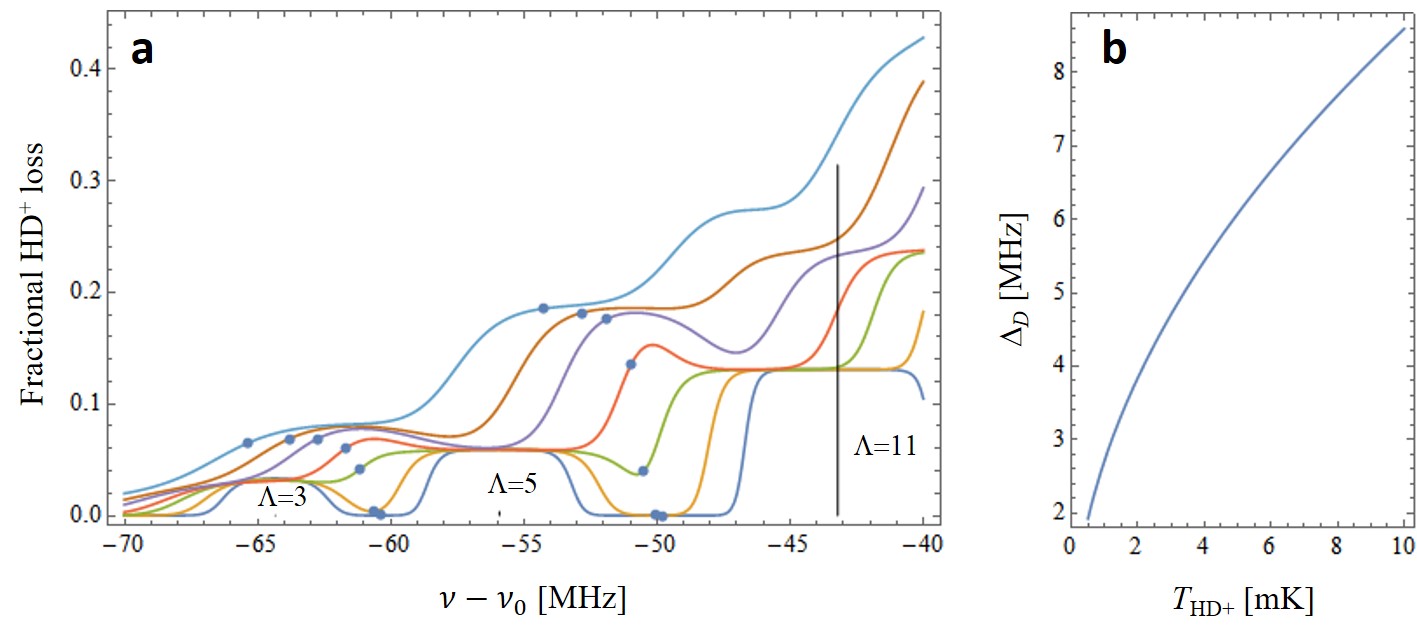}
\caption{(a) REMPD signal plotted against $\nu-\nu_0$ for HD$^+$ ion temperatures (from top to bottom) of 10~mK, 7~mK, 5~mK, 3~mK, 2~mK, 1~mK and 0.5~mK, respectively. The curves are obtained for a BBR temperature of 5~K, which effectively suppresses the population redistribution between rotational states in $v = 0$ vibrational level. For each ion temperature, the data points indicate the values of $\nu_X$ computed for two pairs of hyperfine transitions, namely those involving the transitions at $-64.3$~MHz and $-55.9$~MHz, and the transitions at $-55.9$~MHz and $-43.2$~MHz. Ghost features occur for HD$^+$ temperatures in excess of 2~mK, and shift as the Doppler width increases. (b) Doppler width $\Delta_D$ of the $(v,L):(0,3)\rightarrow(4,2)$ transition at 1442~nm in HD$^+$ in the temperature range considered here.}
\label{fig:5KBBR}
\end{figure}

The qualitative picture of Fig.~\ref{fig:qualplot}(b) ascribes the appearance of ghost features to a combination of saturation and partial overlap of Doppler-broadened lines. However, our rate equation model also includes redistribution of population by BBR over various rotational states within the $v=0$ manifold. To assess the effect of BBR on ghost features, we have reproduced Fig.~\ref{fig:temp}(a) for a BBR temperature of 5~K while leaving all other parameters unchanged; see Fig.~\ref{fig:5KBBR}(a). At 5~K BBR temperature, only the $v=0, L=0$ rotational state of HD$^+$ is populated, thereby effectively suppressing the redistribution of population to the $v=0, L=3$ lower state of the transition. Note that to produce a REMPD signal at 5~K, an appreciable initial population in $v=0, L=3$ is required. Therefore we have calculated the spectra of Fig.~\ref{fig:5KBBR}(a) assuming all population initially being in $v=0, L=3$. Figure~\ref{fig:5KBBR}(a) reveals that ghost features appear also in the absence of population redistribution by BBR, and the overall shape of the spectra resembles that of the spectra at $T_\text{BBR}=300$~K of Fig.~\ref{fig:temp}(a).

\begin{figure}[h!]
\centering
\includegraphics[width=1.0\linewidth]{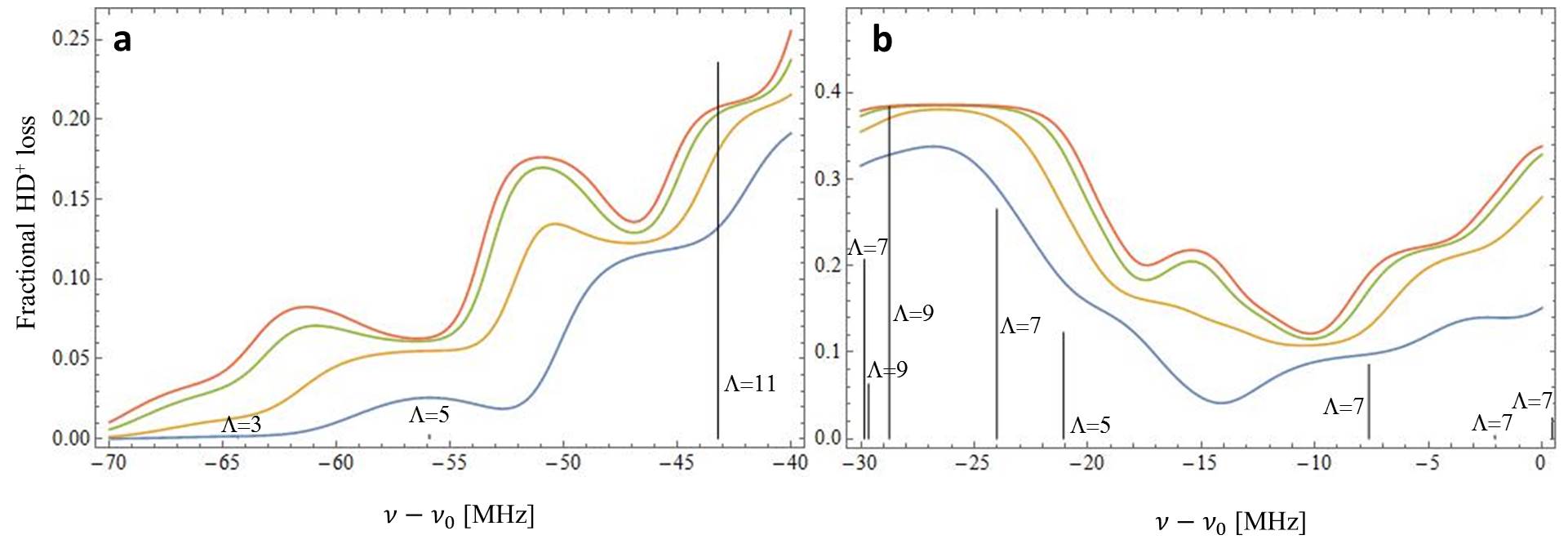}
\caption{(a,b) REMPD signal plotted against $\nu-\nu_0$ for different laser intensities and an ion temperature of 5~mK. From top to bottom, plots correspond to laser powers of 10~mW, 5~mW, 1~mW and 100~$\mu$W, respectively, focused in a 100~$\mu$m Gaussian beam waist. The hyperfine components are indicated by black sticks. The position of one of the ghost features in Fig.~\ref{fig:intensity}(a) appears to shift from $-48$~MHz at 100~$\mu$W to $-51$~MHz at 10~mW. The hyperfine line at $-55.9$~MHz undergoes an even larger apparent shift: if the power increases from 100~$\mu$W to 10~mW, the peak moves from $-55.9$~MHz to $-62$~MHz, meanwhile transforming from a true hyperfine feature into a ghost feature. Also in Fig.~\ref{fig:intensity}(b), a ghost feature near $-15$~MHz becomes visible as the laser intensity is increased.}
\label{fig:intensity}
\end{figure}

Also the intensity of the probe laser has some effect on the strength and position of the ghost features. In Fig.~\ref{fig:intensity}, the signal is plotted for different probe laser intensities. It can be seen that at relatively low laser intensity (100~$\mu$W in a $100~\mu$m beam waist), the peaks are weak. If the laser power is increased to $\geq1$~mW so as to dissociate the excited molecules more efficiently (and thereby induce more saturation), the ghost features not only increase in strength but may also shift spectrally [towards more negative frequency in Fig.~\ref{fig:intensity}(a)]. As the laser intensity is increased further, the shifting of these features becomes smaller. Figure~\ref{fig:intensity}(a) also illustrates that random laser intensity variations may translate to additional signal noise with a spread that is enhanced at the location of a ghost feature, and reduced in between the ghost feature and a true hyperfine line [for example near $-46.5$~MHz in Fig.~\ref{fig:intensity}(a)].

\begin{figure}[!ht]
\centering
\includegraphics[width=1.0\linewidth]{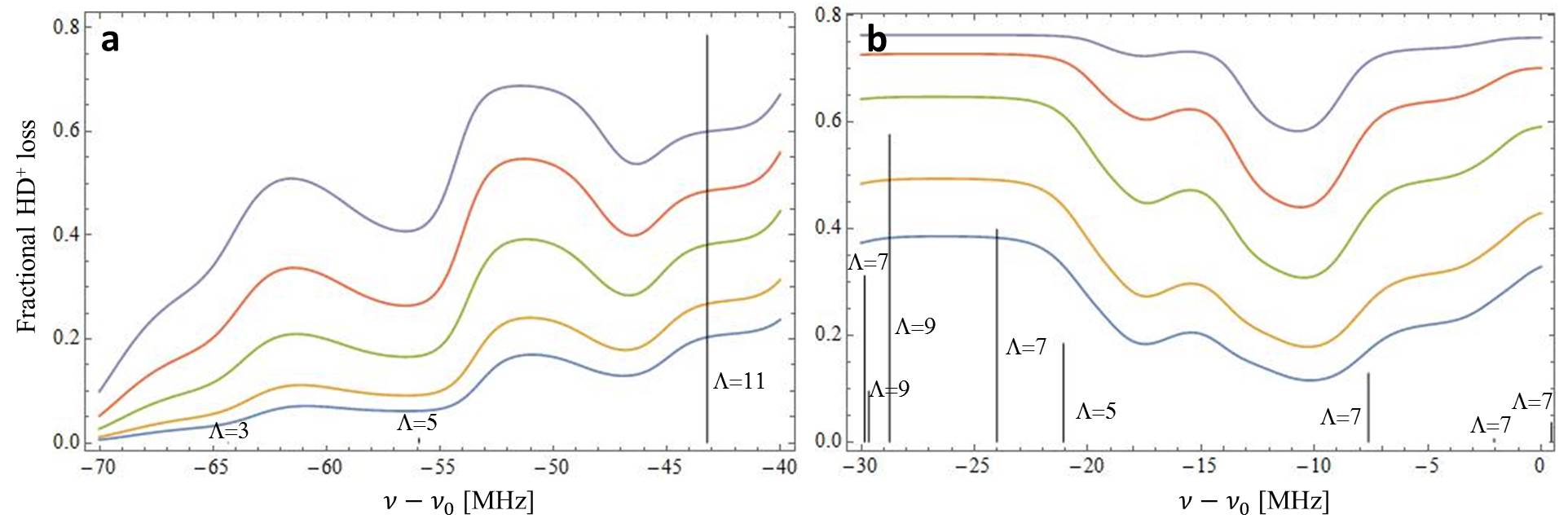}
\caption{REMPD signal plotted against $\nu-\nu_0$ for different probe times, 5~mK ion temperature and 5~mW laser power focused in a 100~$\mu$m waist. From top to bottom, plots correspond to 200~s, 100~s, 50~s, 20~s and 10~s probe times, respectively. Ghost features at $-51$~MHz and $-62$~MHz in Fig.~\ref{fig:probetime}(a) and $-15$~MHz in Fig.~\ref{fig:probetime}(b) become more prominent for longer probe durations, until the duration becomes long enough to nearly deplete the entire initial population, and spectral flattening occurs.}
\label{fig:probetime}
\end{figure}

Another parameter that may affect the ghost features is the time for which the transition is probed. The
longer %($\geq50$ s probe time)
the transitions are probed and the individual transitions get saturated, the more prominent these features become (Fig.~\ref{fig:probetime}), until the duration becomes long enough to nearly deplete the entire initial population and spectral flattening occurs [Fig.~\ref{fig:probetime}(b)]. This can be attributed to the redistribution of population among rotational levels in the $v = 0$ manifold, which initially replenishes the population of the lower states of the transition (thereby limiting saturation), but on longer timescales leads to depletion of the sample of HD$^+$ ions and, thus, to enhanced saturation. Below we will show that for spectroscopy of molecular ions with a permanent electric dipole moment (such as HD$^+$), the appearance of ghost features is under certain conditions not determined by the probe duration, but by the lifetime of the initial state under the influence of BBR-driven rotational transitions and spontaneous decay.

The ghost features reported here bear some resemblance with the cross-over features seen in saturated absorption spectroscopy. However, the origin and nature of these ghost features are completely different. Ghost peaks arise from interaction with a single photon only, whereas cross-over features in saturated absorption spectroscopy involve two counterpropagating photons. Also, unlike the cross-over resonances in saturated absorption scheme, the ghost features reported here do not lie exactly halfway between two transitions. Moreover, their positions depend on probe laser intensity, saturation, and the temperature of the HD$^+$ ions. In the remainder of this Section, we explain the frequency shift of ghost features with changes in the HD$^+$ temperature and Doppler width, and we quantify the conditions under which ghost features may occur.

Let us consider two hyperfine transitions with center frequencies $\nu_1$ and $\nu_2$, coupling lower hyperfine levels $1,2$ to excited states $1',2'$ respectively. We assume that neither transitions from state $1$ to state $2'$ nor from $2$ to $1'$ are allowed, and that there are no other transitions nearby. In this case we can denote the corresponding absorption rate coefficients by $B_1$ and $B_2$. If we consider only levels 1 and 2 and ignore interaction with BBR, then it follows from Eqs.~(\ref{eq:rateeq}) and (\ref{eq:Mrempd}) that the interaction of the probe laser (having frequency $\nu$) with the two lines leads to a REMPD rate proportional to
\begin{equation}
- \sum_{i=1}^{2}\Lambda_i B_{i}D_z \left( 2\pi (\nu -\nu_i) ,T_{\text{HD}^+} \right) \frac{I_\text{probe}}{c}.
\label{eq:gaussiansum}
\end{equation}
As before, we assume that any excited-state population is dissociated with unit probability. We can study the  appearance and behavior of ghost features in between these two hyperfine components by considering the saturation levels near the point where the two lineshapes in Eq.~(\ref{eq:gaussiansum}) intersect as follows. The two terms in Eq.~(\ref{eq:gaussiansum}) correspond to Gaussian lineshape functions centered at $\nu_1$ and $\nu_2$, respectively, both having a full-width-half-maximum Doppler width $\Delta_D$, for which the intersection occurs at a frequency $\nu_X$ given by
\begin{equation}
\nu_X=\frac{\nu_1+ \nu_2}{2} - \frac{\Delta_D^2}{8 (\nu_1-\nu_2)\ln{2}} \ln{ \frac{\Lambda_1 B_{1}}{\Lambda_2 B_{2}}}.
\label{eq:nuX}
\end{equation}
Note that $\nu_X$, through the Doppler width $\Delta_D=\left(8 \ln{2}\, k_B T_{\text{HD}^+} /m\right)^{1/2}/\lambda$, depends on $T_{\text{HD}^+}$ (here $k_B$ is Boltzmann's constant and $\lambda=c/\nu)$. We furthermore point out that Eq.~(\ref{eq:nuX}) requires only relative linestrengths. In Fig.~\ref{fig:5KBBR}(a), we show the positions $\nu_X$ for two pairs of transitions and for various HD$^+$ temperatures. While the peak position of a ghost feature often is displaced slightly from $\nu_X$ (depending on the relative strengths and lower-state multiplicities of the hyperfine transitions involved), it is typically found in the vicinity of $\nu_X$. Equation~(\ref{eq:nuX}) therefore also gives some quantitative insight in the shift of the ghost feature frequency with Doppler width and transition strength.% (provided that the frequencies and relative transition strengths of both transitions are known).

Knowing the value of $\nu_X$, we can also estimate the conditions under which ghost features may occur. By definition, at $\nu=\nu_X$, the populations in levels 1 and 2 are dissociated at the same rate. Therefore, with $n(t)$ being the sum of the populations in levels 1 and 2, the population in each level $i=1,2$ is given by $n(t) \Lambda_i/(\Lambda_1+\Lambda_2)$, and $n(t)$ can be described by as a single exponential decay as follows,
\begin{equation}
n(t)= n(0) \exp{\left( -t \frac{2\Lambda_1}{\Lambda_1+\Lambda_2} B_{1}D_z \left( 2\pi \left( \nu_X -\nu_1 \right) ,T_{\text{HD}^+} \right) \frac{I_\text{probe}}{c} \right) },
\label{eq:expdecay}
\end{equation}
again emphasizing that $\nu_X $ is a function of $T_{\text{HD}^+}$. Replacing $t$ with the total interaction time, $\tau$, reveals whether significant saturation due to depletion occurs (\textit{i.e.} $n(\tau)/n(0)\ll 1$). In the absence of spontaneous decay of the initial state or interaction with BBR, $\tau$ is simply the total probe duration. However, in the case of molecular ions with a permanent electric dipole moment (such as HD$^+$), the value of $\tau$ also depends on the lifetime of the initial state under the influence of BBR-induced rotational transitions and spontaneous decay. The above example of an ensemble of HD$^+$, initially being prepared in the $v=0, L=3$ rotational state and interacting with 5~K BBR and the REMPD lasers, is illustrative in this respect. In this case, the initial $v=0, L=3$ state has a spontaneous lifetime of 4.4~s, which is considerably shorter than the 10~s probe duration used in most calculations in this work. Let us consider the hyperfine transitions located at $\nu-\nu_0=-55.9$~MHz and $\nu-\nu_0=-43.2$~MHz in Fig.~\ref{fig:5KBBR}(a) (which we label 1 and 2, respectively), having absorption rate coefficients $B_{1}=6.4\times 10^{11}$~m$^3$(rad/s)/(J\,s) and $B_{2}=2.5\times 10^{13}$~m$^3$(rad/s)/(J\,s), and multiplicities $\Lambda_1=5$ and $\Lambda_2=11$. Inserting these values into Eq.~(\ref{eq:nuX}) and combining the result with Eq.~(\ref{eq:expdecay}), we find for $\tau=4.4$~s that $n(\tau)/n(0) \approx 1$ for $T_{\text{HD}^+}<1.3$~mK (\textit{i.e.} no saturation and therefore no ghost features), while $n(\tau)/n(0) \approx 0$ for $T_{\text{HD}^+}>2.3$~mK, indicating strong saturation so that ghost features may be expected (see also Fig.~\ref{fig:fig9}). These findings, obtained from the relatively simple Eqs.~(\ref{eq:nuX}) and (\ref{eq:expdecay}), is in agreement with the curves shown in Fig.~\ref{fig:5KBBR}(a) obtained from the full model. We also find that for intermediate values, $0.05 \lesssim n(\tau)/n(0) \lesssim 0.95$, individual hyperfine components start to overlap significantly, but ghost features typically remain absent.
\begin{figure}[!ht]
\centering
\includegraphics[width=0.8\linewidth]{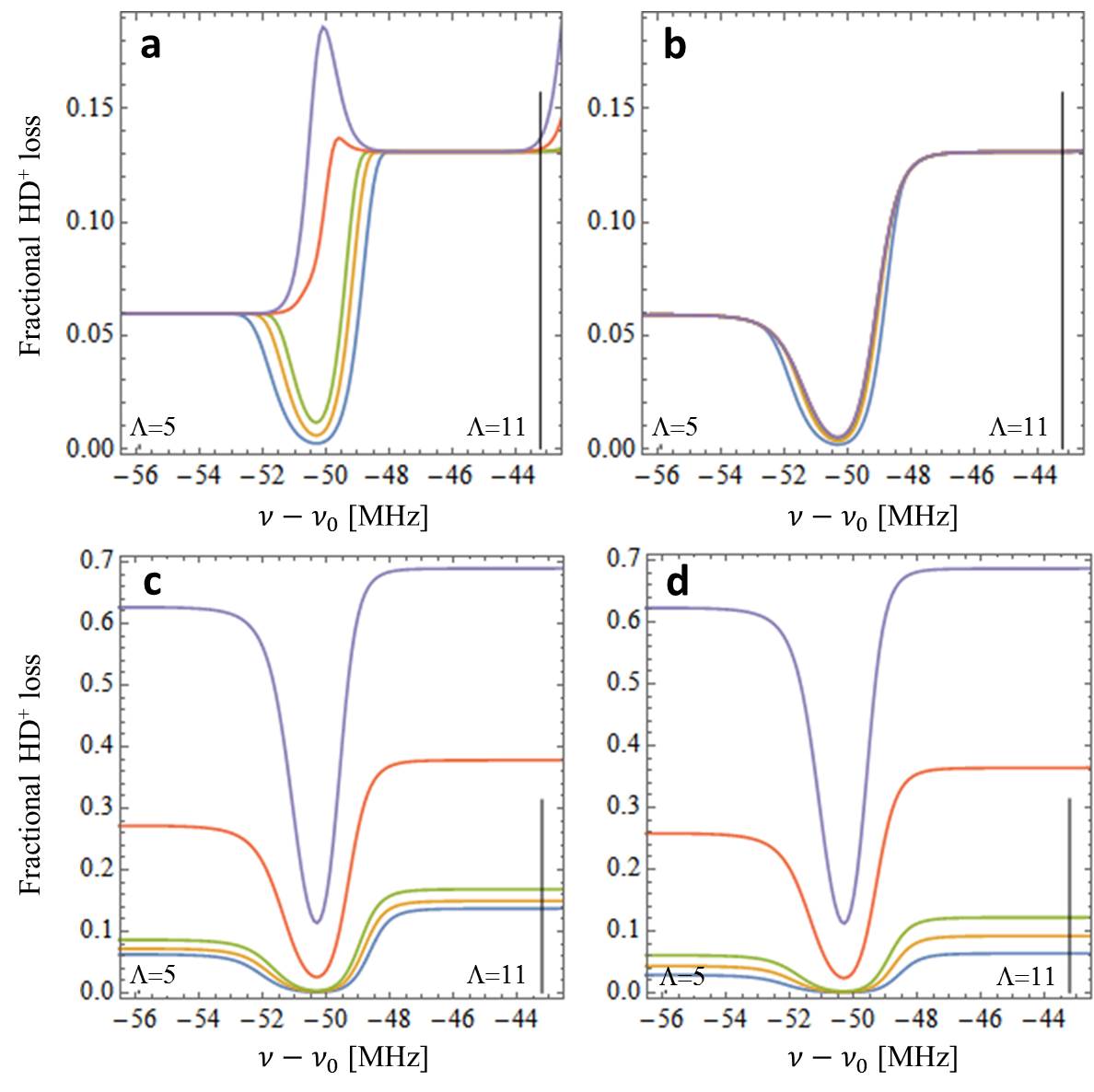}
\caption{REMPD signal plotted against $\nu-\nu_0$ at 5~mK ion temperature and 5~mW laser power focused in a 100~$\mu$m waist, for varying probe time (from top to bottom, 500~s, 100~s, 10~s, 5~s, and 2~s, respectively) and for four different scenarios. (a) Initial population in $v=0, L=3$ only. No spontaneous decay of the initial state and no interaction with BBR included. (b)  Initial population in $v=0, L=3$ only. Spontaneous decay from $v=0,L=3$ to $v=0,L=2$ allowed, but population redistribution by BBR is not included. (c)  Initial population in $v=0, L=3$ only. Spontaneous decay and interaction with 300-K BBR are included. (d) Initial population corresponding to a thermal distribution at 300~K (covering $v=0, L=0-5$), and spontaneous decay and interaction with 300-K BBR are included.}
\label{fig:fourscenarios}
\end{figure}

To provide more insight into the role of spontaneous decay and BBR-induced rotational transitions involving the $v=0,L=3$ initial state in relation to Eq.~(\ref{eq:expdecay}), we have plotted the same part of the spectrum, for various probe times, for four different scenarios in Fig.~\ref{fig:fourscenarios}. A relatively low HD$^+$ temperature of 1.5~mK is used for all four scenarios. In Fig.~\ref{fig:fourscenarios}(a), we fully disable spontaneous decay and BBR by removing the term $M_\text{BBR}\cdot \rho(t)$ from Eq.~(\ref{eq:rateeq}). As a consequence, $\tau$ is determined entirely by the probe duration. We note that this scenario is similar to laser spectroscopy of trapped molecular ions without a permanent electric dipole moment, such as H$_2^+$ or N$_2^+$. For shorter probe times the two hyperfine components at $-55.9$~MHz and $-43.2$~MHz are resolved, but for longer probe durations ($\geq 100$~s) a ghost feature appears. This is fully in line with Eq.~(\ref{eq:expdecay}), which predicts that ghost features appear at 1.5~mK for probe times of 100~s and longer (see Fig.~\ref{fig:fig9}). This situation changes drastically when spontaneous emission from the $v=0,L=3$ initial state to $v=0,L=2$ is introduced, as shown in Fig.~\ref{fig:fourscenarios}(b). In this case, the interaction time $\tau$ is effectively limited to the 4.4~s lifetime of the $v=0,L=3$ state, and probing for 10~s or longer does not change the spectrum. In Fig.~\ref{fig:fourscenarios}(c) we again solve the rate equations starting from an initial population in $v=0, L=3$ only [as was also done for Figs.~\ref{fig:fourscenarios}(a) and (b)], but now the HD$^+$ ensemble is allowed to fully interact with the 300-K BBR and to undergo spontaneous emission. In this scenario, spontaneous emission initially leads to depletion of $v=0,L=3$, but this is counteracted by BBR-induced transitions back from $v=0,L=2$ to $v=0,L=3$. This population redistribution occurs independently of the frequency $\nu$ of the probe laser, and therefore it merely leads to an overall vertical scaling of the spectral signal towards higher signal levels (and not to the formation of ghost features). A similar vertical scaling is visible in Fig.~\ref{fig:probetime}. This behavior suggests that for $\tau$ in Eq.(\ref{eq:expdecay}) we must use the typical timescale at which BBR-induced depletion and population redistribution occurs, if the goal is to predict whether ghost features appear or not. This timescale can be computed by summing all the depopulation rates (BBR-induced and spontaneous) from the $v=0, L=3$ level and taking the inverse, yielding $\tau=0.96$~s at 300~K. Figure~\ref{fig:fourscenarios}(d) is similar to Fig.~\ref{fig:fourscenarios}(c), with the only difference being that here the initial rotational distribution corresponds to a thermal distribution at 300~K, so that also states with $v=0, L=0,1,2,4,5$ are populated significantly. As a consequence, the redistribution to $v=0,L=3$ takes longer, which manifests itself as a smaller overall increase of the signal with increasing probe time. Nevertheless, in both scenarios of Fig.~\ref{fig:fourscenarios}(c) and (d), the signals at long probe times (500~s and longer) saturate at nearly the same level. Also for the conditions used to obtain Fig.~\ref{fig:fourscenarios}(d), we must use $\tau=0.96$~s, and in Fig.~\ref{fig:fig9} we plot $n(\tau=0.96~\text{s} )/n(0)$, predicting well resolved lines for $T_{\text{HD}^+}<1.5$~mK, and ghost features for $T_{\text{HD}^+}>2.8$~mK, in good agreement with the curves of Fig.~\ref{fig:temp} and Fig.~\ref{fig:fourscenarios}(d).
\begin{figure}
\centering
\includegraphics[width=0.5\linewidth]{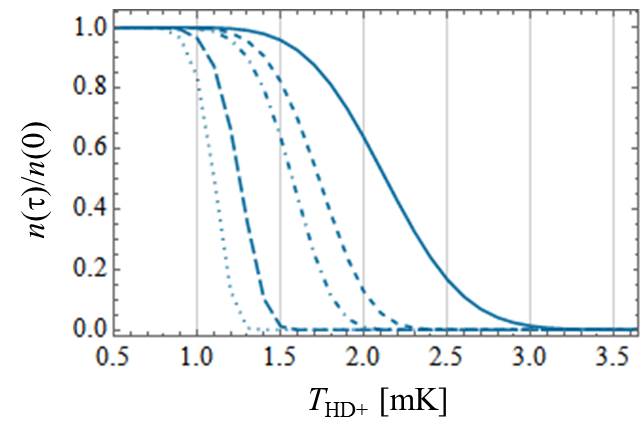}
\caption{Value of $n(\tau )/n(0)$ versus $T_{\text{HD}^+}$ as given by Eq.~(\ref{eq:expdecay}), for $\tau=0.96$~s (solid curve), $\tau=4.4$~s (dashed curve), $\tau=10$~s (dot-dashed curve), $\tau=100$~s (long-dashed curved), and $\tau=500$~s (dotted curve). A value of $n(\tau )/n(0) \approx 1$ implies that individual hyperfine components are well resolved and ghost features are absent, while $n(\tau )/n(0) \approx 0$ indicates that ghost features may be expected to occur. Intermediate values ($0.05 \lesssim n(\tau)/n(0) \lesssim 0.95$) typically imply that hyperfine components start to overlap significantly, while ghost features remain absent.}
\label{fig:fig9}
\end{figure}
\section{Significance of ghost features for spectroscopy of trapped ions} \label{sec:impact}
Obviously, the ghost features reported here may lead to false line identifications if the observed spectra are not analyzed with an adequate lineshape model. We emphasize that this phenomenon is not limited to trapped molecular ions: in principle any system, atomic or molecular, with partly overlapping hyperfine or fine structure may exhibit ghost features in the case of strong saturation and resulting depletion of the sample. The possible influence of ghost features should also be observed in two-photon spectroscopic schemes such as that proposed by Tran \textit{et al.}~\cite{Tran2013}, which involves quasi-degenerate photons inducing near-resonant transitions between $(v,L):(0,3)\rightarrow(4,2)$ and $(v,L):(4,2)\rightarrow(9,3)$. As pointed out by Tran \textit{et al.}, the REMPD signal in this case will consist of a Doppler-free two-photon contribution on top of a Doppler-broadened background. To suppress the latter, it is important to tune the first laser in the wing of the single-photon $(v,L):(0,3)\rightarrow(4,2)$ spectrum, more than one Doppler width away from each hyperfine component~\cite{Tran2013}. Turning now to Fig.~\ref{fig:temp}(a), we note that the hyperfine sticks at $-56$~MHz and $-43$~MHz are 13~MHz apart, which is more than twice the 6~MHz Doppler width at 5~mK. One might therefore naively guess that the Doppler-broadened signal should be suppressed near $-49.5$~MHz, and that this might be a suitable frequency for the first laser in the two-photon spectroscopy to be tuned to. However, at 5~mK this region of the spectrum exhibits an \textit{enhanced} Doppler-broadened signal in the form of a ghost feature. In Ref.~\cite{Karr2016b} we identified suitable positions in the $(v,L):(0,3)\rightarrow(4,2)$ spectrum for the two-photon spectroscopy envisaged by Tran \textit{et al.}~\cite{Tran2013}.

\section{Summary and conclusion} \label{sec:summary}
In this article, we have used the rate equation model developed by Biesheuvel \textit{et al.} in Ref.~\cite{Biesheuvel2016} to study the $(v,L):(0,3)\rightarrow(4,2)$ single-photon transition in a finite-sized sample of HD$^+$ molecular ions. We analyze and explain, for the first time to our knowledge, the appearance of ghost features in the Doppler-broadened hyperfine structure of the above transition. Our analysis reveals a dependence of ghost feature prominence and position on the Doppler width, relative transition strengths, and saturation levels, and we quantify the conditions under which ghost features may appear, as well as their frequency dependence on Doppler width and relative linestrengths. We find that ghost features may manifest themselves in rovibrational spectra of molecular ions with and without a permanent electric dipole moment (which exhibit interaction and no interaction with BBR, respectively), albeit under different conditions. We conclude that knowledge of these spectral features is important for assigning lines in any precision spectroscopy of finite-sized samples of atoms in which a significant fraction of the sample becomes depleted, and where the values of the transition frequencies are not known accurately enough \textit{a priori}, or the samples not cold enough to rule out overlap of the Doppler-broadened line-profiles of adjacent transitions. Our conclusion not only pertains to composite lineshapes in vibrational spectroscopy of cold trapped ions in the optical domain, but to any Doppler-broadened and strongly saturated spectroscopy of finite samples of trapped particles, including laser-induced transitions between atomic or molecular electronic states, and rotational spectroscopy of molecular ions at THz frequencies. We also point out that our results may be extended with no restrictions to other (higher) temperature domains. This highlights the general importance of a rate-equation model to properly understand and interpret the data obtained from precision spectroscopy with finite-sized samples having a rich structure with overlapping line profiles.
%We argue that these kind of features are common in spectroscopy on finite sized samples with partly overlapping line-profiles and are very important in interpreting the data from these experimental studies.

\section*{Acknowledgements}
This research was funded through the Netherlands Foundation for Fundamental Research on Matter (FOM). J.C.J.K. thanks the Netherlands Organisation for Scientific Research (NWO) and the Netherlands Technology Foundation (STW) for support.

%\nocite{*}
\bibliographystyle{elsarticle-num}
\bibliography{bibliographyJMS}% Produces the bibliography via BibTeX.

\begin{thebibliography}{10}
\expandafter\ifx\csname url\endcsname\relax
  \def\url#1{\texttt{#1}}\fi
\expandafter\ifx\csname urlprefix\endcsname\relax\def\urlprefix{URL }\fi
\expandafter\ifx\csname href\endcsname\relax
  \def\href#1#2{#2} \def\path#1{#1}\fi

\bibitem{Biesheuvel2016}
J.~Biesheuvel, J.-{\relax Ph}. Karr, Karr, L.~Hilico, K.~S.~E. Eikema,
  W.~Ubachs, J.~C.~J. Koelemeij, Probing {QED} and fundamental constants
  through laser spectroscopy of vibrational transitions in {HD}$^+$, Nat.
  Commun. 7~(10385).
\newblock \href {http://dx.doi.org/10.1038/ncomms10385}
  {\path{doi:10.1038/ncomms10385}}.

\bibitem{Korobov2014a}
V.~I. Korobov, L.~Hilico, J.-{\relax Ph}. Karr, $m \alpha^7$-order corrections
  in the hydrogen molecular ions and antiprotonic helium, Phys. Rev. Lett. 112
  (2014) 103003.
\newblock \href {http://dx.doi.org/10.1103/PhysRevLett.112.103003}
  {\path{doi:10.1103/PhysRevLett.112.103003}}.

\bibitem{Korobov2014b}
V.~I. Korobov, L.~Hilico, J.-{\relax Ph}. Karr, Theoretical transition
  frequencies beyond 0.1 ppb accuracy in {H}$_{2}^{+}$, {H}{D}$^{+}$, and
  antiprotonic helium, Phys. Rev. A 89 (2014) 032511.
\newblock \href {http://dx.doi.org/10.1103/PhysRevA.89.032511}
  {\path{doi:10.1103/PhysRevA.89.032511}}.

\bibitem{Bakalov2006}
D.~Bakalov, V.~I. Korobov, S.~Schiller, Phys. Rev. Lett. 97 (2006) 243001.

\bibitem{Korobov2016}
V.~I. Korobov, J.~C.~J. Koelemeij, L.~Hilico, J.-{\relax Ph}. Karr, Theoretical
  hyperfine structure of the molecular hydrogen ion at the 1 ppm level, Phys.
  Rev. Lett. 116~(053003).
\newblock \href {http://dx.doi.org/10.1103/PhysRevLett.116.053003}
  {\path{doi:10.1103/PhysRevLett.116.053003}}.

\bibitem{Roth2006}
B.~Roth, J.~C.~J. Koelemeij, H.~Daerr, S.~Schiller, Rovibrational spectroscopy
  of trapped molecular hydrogen ions at millikelvin temperatures, Phys. Rev. A
  74 (2006) 040501.
\newblock \href {http://dx.doi.org/10.1103/PhysRevA.74.040501}
  {\path{doi:10.1103/PhysRevA.74.040501}}.

\bibitem{Koelemeij2007b}
J.~C.~J. Koelemeij, B.~Roth, S.~Schiller, Blackbody thermometry with cold
  molecular ions and application to ion-based frequency standards, Phys. Rev. A
  76 (2007) 023413.

\bibitem{Bressel2012}
U.~Bressel, A.~Borodin, J.~Shen, M.~Hansen, I.~Ernsting, S.~Schiller,
  Manipulation of individual hyperfine states in cold trapped molecular ions
  and application to {H}{D}$^+$ frequency metrology, Phys. Rev. Lett. 108
  (2012) 183003.
\newblock \href {http://dx.doi.org/10.1103/PhysRevLett.108.183003}
  {\path{doi:10.1103/PhysRevLett.108.183003}}.

\bibitem{Zhong2015}
Z.-{\relax X}. Zhong, X.~Tong, Z.-{\relax C}. Yan, T.-{\relax Y}. Shi,
  High-precision spectroscopy of hydrogen molecular ions, Chin. Phys. B 24
  (2015) 053102.
\newblock \href {http://dx.doi.org/10.1088/1674-1056/24/5/053102}
  {\path{doi:10.1088/1674-1056/24/5/053102}}.

\bibitem{Blythe2005}
P.~Blythe, B.~Roth, U.~Fr\"ohlich, H.~Wenz, S.~Schiller, Production of
  ultracold trapped molecular hydrogen ions, Phys. Rev. Lett. 95 (2005) 183002.
\newblock \href {http://dx.doi.org/10.1103/PhysRevLett.95.183002}
  {\path{doi:10.1103/PhysRevLett.95.183002}}.

\bibitem{SpezeskiThesis}
J. J. Spezeski, Ph.D. thesis, Yale University (1977).

\bibitem{Koelemeij2007a}
J.~C.~J. Koelemeij, B.~Roth, A.~Wicht, I.~Ernsting, S.~Schiller, Vibrational
  spectroscopy of {H}{D}$^+$ with 2-ppb accuracy, Phys. Rev. Lett. 98 (2007)
  173002.

\bibitem{Tran2013}
V.~Q. Tran, J.-{\relax Ph}. Karr, A.~Douillet, J.~C.~J. Koelemeij, L.~Hilico,
  Two-photon spectroscopy of trapped {H}{D}$^+$ ions in the {L}amb-{D}icke
  regime, Phys. Rev. A 88 (2013) 033421.
\newblock \href {http://dx.doi.org/10.1103/PhysRevA.88.033421}
  {\path{doi:10.1103/PhysRevA.88.033421}}.

\bibitem{Karr2016}
J.-{\relax Ph}. Karr, L.~Hilico, J.~C.~J. Koelemeij, V.~I. Korobov, Hydrogen
  molecular ions for improved determination of fundamental constants,
  arXiv:1605.05456 [physics.atom-ph].

\bibitem{Koelemeij2011}
J.~C.~J. Koelemeij, Infrared dynamic polarizability of {H}{D}$^+$ rovibrational
  states, Phys. Chem. Chem. Phys. 13~(42) (2011) 18844--18851.

\bibitem{Bakalov2011}
D.~Bakalov, V.~I. Korobov, S.~Schiller, J. Phys. B: At. Mol. Opt. Phys. 44
  (2011) 025003.

\bibitem{Biesheuvel2016b}
J. Biesheuvel \textit{et al.}, in preparation.

\bibitem{Karr2016b}
J.-{\relax Ph}. Karr, S.~Patra, J.~C.~J. Koelemeij, J.~Heinrich, N.~Silitoe,
  A.~Douillet, L.~Hilico,
  \href{http://stacks.iop.org/1742-6596/723/i=1/a=012048}{Hydrogen molecular
  ions: new schemes for metrology and fundamental physics tests}, Journal of
  Physics: Conference Series 723~(012048).
\newline\urlprefix\url{http://stacks.iop.org/1742-6596/723/i=1/a=012048}

\end{thebibliography}

\end{document}